\documentclass{moriond}
\usepackage{booktabs}
\usepackage{colortbl}
\usepackage{xcolor}
\usepackage{amsmath}
\usepackage{caption}
\usepackage{cite} 

\usepackage{subcaption}



\usepackage{float}          

\DeclareUnicodeCharacter{0301}{\'{}}   

\def\Journal#1#2#3#4{{#1} {\bf #2}, #3 (#4)}

\def\be{\begin{equation}}
\def\ee{\end{equation}}
\def\bea{\begin{eqnarray}}
\def\eea{\end{eqnarray}}

\definecolor{lightgray}{rgb}{0.83, 0.83, 0.83}
\definecolor{myred}{RGB}{255, 0, 0} 

\newcommand{\Photo}{} 

\begin{document}
\vspace*{4cm}
\title{Performance of the advanced gamma-ray trigger system for the High Energy Cosmic Radiation Detection (HERD) facility}

\author{ Keerthana Rajan Lathika on behalf of the HERD Collaboration}

\address{Institut de Fisica d’Altes Energies (IFAE), BIST, Barcelona, Spain}

\maketitle\abstracts{
The High Energy Cosmic-Radiation Detection (HERD) facility has been proposed as a leading experiment on China’s Space Station(CSS). Scheduled for installation around 2027, HERD is expected to operate for at least a decade. The main scientific objectives include indirect detection of dark matter with unprecedented sensitivity, studying the cosmic-ray spectrum and composition up to the knee, and observing all-sky gamma rays with energies above 100 MeV. HERD is designed as a large-acceptance telescope with a unique design aimed at maximizing its efficiency. It comprises a central 3D imaging calorimeter (CALO) made of LYSO crystals, encircled by four complementary subdetectors on its top and four lateral faces: the scintillating fiber tracking detector (FIT), the plastic scintillator detector (PSD), the silicon charge detector (SCD), and a transition radiation detector (TRD) on one lateral side. To fully harness HERD gamma-ray detection capabilities down to 100 MeV, an advanced ultra-low-energy gamma-ray (ULEG) trigger system has been developed. We present an extensive overview of the design, performance, and optimization of the gamma-ray trigger system supported by software simulations and preliminary results from the successful implementation of the HERD prototype at CERN’s PS and SPS beam test campaigns in Fall 2023.
}

\section{Introduction: ULEG trigger}

HERD's ultra-low-energy gamma-ray (ULEG) trigger system, operating at three trigger levels (L0, L1, and L2), integrates information from FIT\cite{1}, PSD\cite{2}, and CALO subdetectors to identify specific energy deposition patterns, facilitating precise detection and reconstruction of gamma-ray events to fulfill HERD’s scientific objectives. The function of FIT involves trajectory reconstruction of charged secondaries and direction reconstruction of primary gamma rays. PSD serves as the gamma-ray veto system, and CALO is utilized for energy reconstruction. In its current design, the FIT consists of 5 sectors each corresponding to one of the detector instrumented faces, and each made of 7 tracking planes. Each tracking plane is made of two layers of tracking modules to measure the two orthogonal (X, Y) spatial coordinates. The tracking planes of the top sector are made of 10 + 10 (X+Y) modules, while the tracking planes of the side sectors are made of 6 + 10 (horizontal + vertical) modules. The FIT L0 trigger is based upon the so-called ``three-in-a-row'' (3IR) \cite{3} logic that aims at identifying hits (produced by the passage of charged particles) in at least three consecutive FIT layers, with a maximum displacement of $\Delta M = \pm 1$ FIT modules between consecutive layers. At Level 1, PSD data is integrated with FIT-identified regions of interest, generating PSD-vetoed signals by mapping trajectories that align with the 3IR pattern into the PSD. Finally, CALO energy deposition information is combined at Level 2 (L2).

\section{Performance of the ULEG trigger}

\subsection{Gamma-ray performance}\label{subsec:prod}

Simulations were performed using HerdSoftware, a toolkit based on Geant 4  \cite{4}, to study the gamma-ray performance of the ULEG trigger \cite{5}. Table 1 compares the point spread function (PSF) (68\% \text{ containment fraction)}, the acceptance, and the energy resolution between HERD and Fermi-LAT. The acceptance of HERD is $\sim 5$ times lower than that of Fermi-LAT due to the absence of high-density conversion foils in the tracker. At the same time, this absence improves the PSF. The average diffuse gamma-ray rate is computed as 0.001 counts/s by factoring the CSS orbit, the orientation and lifespan of HERD, and the measured galactic and extragalactic gamma-ray background fluxes by Fermi-LAT. Figure 1 shows the spatial dependence of the measured diffuse gamma-ray flux, in HERD local coordinates. Most of the counts come from the galactic plane, which aligns with the 270\textdegree{} orientation of HERD.

\begin{figure}[h]
\vspace{-0.2cm}
\hspace{-1.5cm} 
\begin{minipage}[b]{0.5\textwidth}
\centering
\includegraphics[width=0.9\linewidth, height=0.6\textwidth, keepaspectratio]{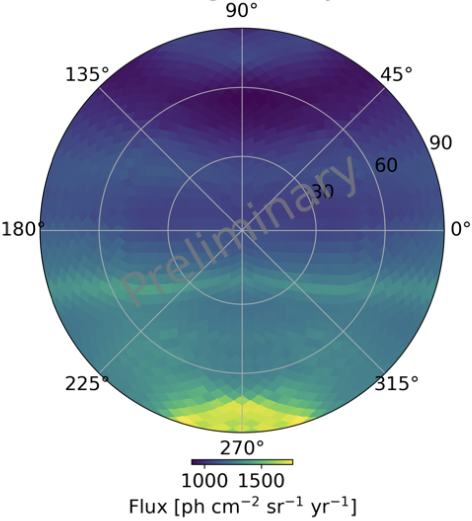}
\captionsetup{justification=centering}
\label{Figure1}
\end{minipage}%
\hspace{0.5cm} 
\begin{minipage}[b]{0.5\textwidth}
\centering
\renewcommand{\arraystretch}{2.4}
\resizebox{\linewidth}{!}{
\fontsize{12}{14}\selectfont 
\begin{tabular}{@{}cccc@{}}
\toprule
\rowcolor{gray!25}
\textbf{Detector} & \textbf{Acceptance} & \textbf{PSF} & \textbf{Energy Resolution} \\
\midrule
\textbf{HERD} & $\sim 0.5 \, \text{m}^{2}\text{sr}$ & $\sim 0.1 \, \text{deg}$ & $\sim 1\%$ \\
& @ $1\text{GeV}$ & @ $10 \, \text{GeV}$ & @ $200 \, \text{GeV}$ \\
\midrule
\textbf{Fermi-LAT} & $\sim 2.5 \, \text{m}^{2}\text{sr}$ & $\sim 0.2 \, \text{deg}$ & $\sim 8\%$ \\
& @ $1\text{GeV}$ & @ $10 \, \text{GeV}$ & @ $200 \, \text{GeV}$ \\
\bottomrule
\end{tabular}
}
\captionsetup{justification=centering}
\label{Table1}
\end{minipage}
\caption{ The figure displays the diffuse gamma-ray flux observed by HERD represented in its local coordinate system $(\theta, \phi)$, where $\theta$ and $\phi$ represent the polar and azimuth angles, respectively. In this system, azimuth angles correspond to directions: $0^\circ$ for forward (aligning with the direction of movement of HERD), $90^\circ$ for port, $180^\circ$ for aft, and $270^\circ$ for starboard. Most of the gamma-ray counts originate from the galactic plane, which lines up with $270^\circ$ azimuth orientation. HERD's $41.5^\circ$ orbit inclination aligns cardinal directions within a $41.5^\circ$ azimuth range depending on its position along the orbit. Table 1: Comparison of acceptance, PSF, and energy resolution between HERD and Fermi-LAT. }
\label{fig:table_and_figure}
\end{figure}

\subsection{ULEG trigger rates for Cosmic ray (CR) particle backgrounds }

In the realm of gamma-ray telescopes, a thorough understanding of instrumental background is indispensable. The PSD is crucial in this context, as its primary role is to provide charged particle background rejection for gamma-ray triggers by acting as a veto system. This function delineates its principal requirement to maintain a high level of charged particle detection efficiency.

CR particles contribute to two distinct instrumental backgrounds. The first one, termed the residual background, arises from the real-world detector design, such as the physical gaps present in the detector. Acceptance of CR particles by the detector is calculated from simulations. A reliable cosmic ray flux model based on experimental data from previous missions such as AMS-01, PAMELA, NINA-2, MARYA-2, and HEPD-Limadou \cite{6} was used to estimate the flux for the geomagnetic latitude range spanned by HERD. The CR background rates at the ULEG L2 level were determined to be approximately 56 counts/s by convolving the flux with the acceptance for different cosmic ray particles (p, He, \( e^- \), and \( e^+ \)), as shown in Figure 2. These results reflect the limitations posed by the physical gaps in the current PSD detector geometry used within the simulations.

The other component of the instrumental background is characterized as the irreducible gamma-ray background \cite{7}, which is produced by charged cosmic rays interacting with the SCD and the passive elements of the payload. Such interactions, including neutral pion decay, bremsstrahlung, and \( e^- \) - \( e^+ \) annihilation produce gamma rays that are indistinguishable from primary ones when passing through the PSD. Integrated rates of the irreducible background have been calculated for different calorimeter energy thresholds for all sectors of HERD, considering a fully hermetic PSD geometry in simulations. The extent of irreducible gamma-ray background is affected by the varying incoming cosmic ray flux at different geomagnetic latitudes, with protons being the primary contributors (92\%) at higher geomagnetic latitudes and positrons (90\%) prevailing at lower latitudes. In Figure 3, the irreducible background rates in the TOP sector of HERD, both at high and low geomagnetic latitudes, are depicted in comparison to the average diffuse gamma-ray background rate. For an energy deposition threshold of 100 MeV in the calorimeter, the integrated rate of irreducible background across all five sectors is calculated to be 0.17 counts/s for high geomagnetic latitude and a rate of 0.05 counts/s for low geomagnetic latitude.  The analysis indicates that the irreducible background constitutes a significant portion of the overall background for HERD. 


\begin{figure}[h]
    \centering
    \begin{minipage}[t]{0.45\textwidth} 
        \centering
        \includegraphics[width=1.1\textwidth, height=0.7\textwidth, keepaspectratio]{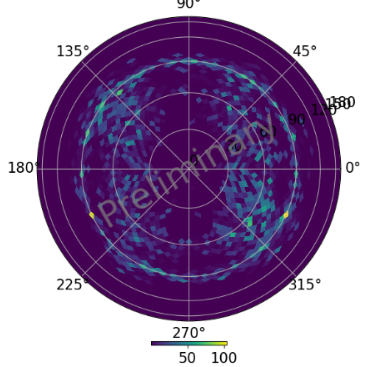}
        \caption{Total CR ULEG-L2 level trigger rate in HERD local coordinates (See Figure 1) as seen from the top. }
    \end{minipage}
    \hspace{0.05\textwidth} 
    \begin{minipage}[t]{0.45\textwidth} 
        \centering
        \includegraphics[width=1.5\textwidth, height=0.7\textwidth, keepaspectratio, trim=0.2cm 0 0 0, clip]{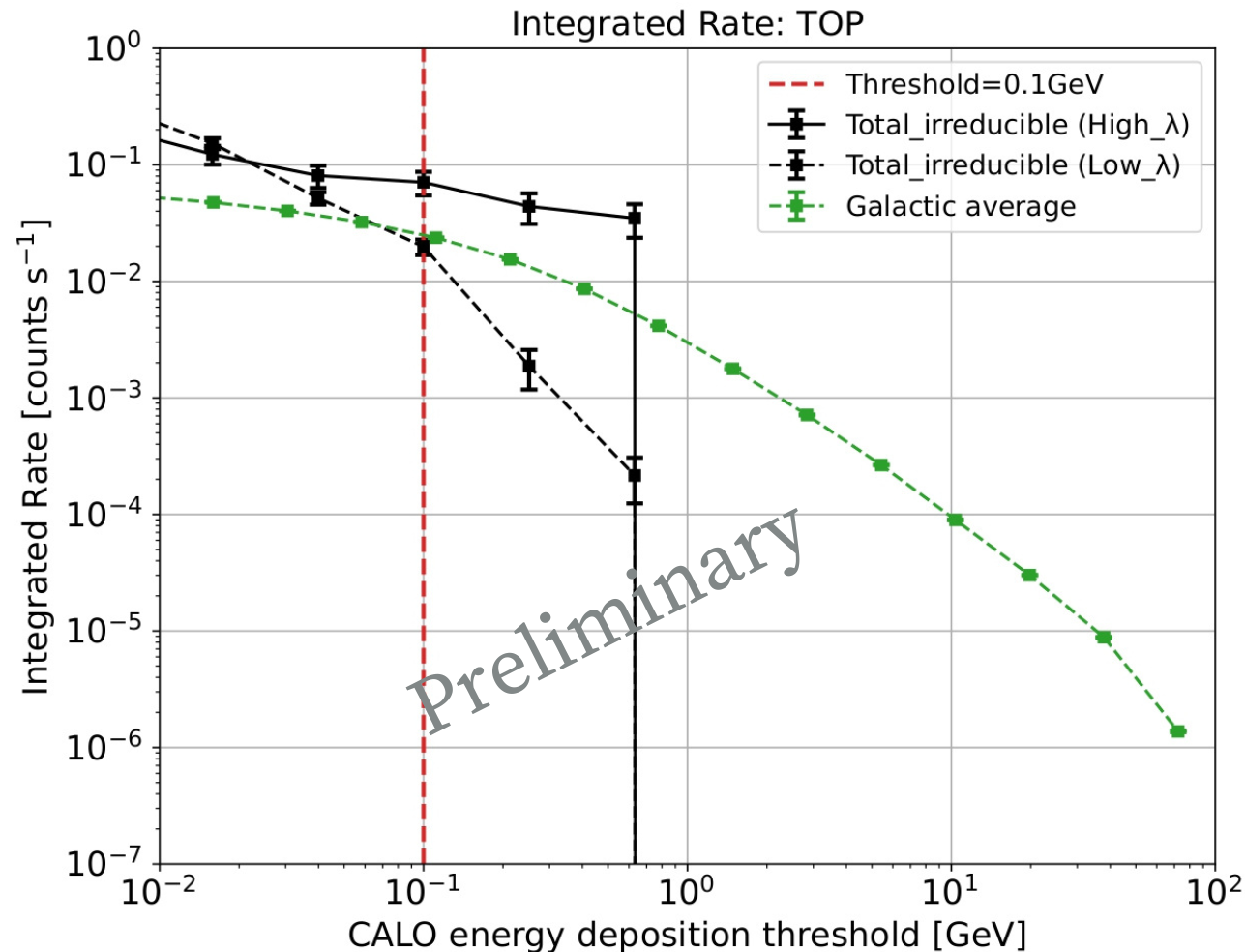}
        \captionsetup{ singlelinecheck=off, width=\linewidth} 
        \captionof{figure}{Irreducible background rates compared to the average diffuse gamma-ray background rate.}
    \end{minipage}
\end{figure}

\section {Test Beam Evaluation of ULEG Trigger Capabilities}

The proof-of-concept prototypes of the PSD and FIT subdetectors of HERD were tested at CERN's PS (Proton Synchrotron) and  SPS (Super Proton Synchrotron) facilities. The main objectives of the beam tests were to integrate the front-end electronics with the main trigger system and data acquisition systems, to investigate the performance of the 16-channel Beta ASIC \cite{8} readout system, to validate the veto trigger in the PSD prototype and to evaluate the 3IR trigger logic using the FIT prototype.
\subsection {Prototype Setup and Test Beam Results for PSD and FIT}
The PSD prototype comprises 8 scintillating bars arranged in two planes with perpendicular orientations. Each bar is equipped with eight SiPMs of two different sizes, optimized for the detection of high and low-charge signals, respectively. The PSD prototype, along with the front-end electronics is shown in Figure 4(left). An internal majority trigger condition is applied, in which the evaluation board distinguishes between signal and noise. This is done by requiring at least 3 out of 4 SiPM of a given size to see a signal above the threshold, to produce the veto signals. The analysis resulted in efficiencies of 99.4\% for the majority of 3/4 triggers and 98.6\% for the majority of 4/4 triggers. During the tests conducted at SPS, the prototype was able to accurately identify nuclei with energies ranging from 250 to 350 GeV. The identified nuclei included helium to iron, as shown in the charge spectrum depicted in Figure 4(right). 

The FIT prototype uses a scaled-down version known as MiniFIT, as shown in Figure 5(a). The MiniFIT is comprised of five layers, each with two FIT modules positioned perpendicular to each other along the horizontal (X) and vertical (Y) axes, respectively. For the beam tests, four layers are instrumented as illustrated in Figure  5(b), each having SiPM channels sensitive to an 8mm wide region, except levels L2 and L3 in the X tower, which have a 16mm width, to effectively test and validate all 3IR trigger signals. The overall sensitive region is 8x8 mm². Figure 5(c) displays the MiniFIT with its front-end electronics, along with other subdetector prototypes at CERN PS. Based on the SPS beam test, considering the straight nuclei tracks in both the X and Y towers, the charge spectrum of nuclei ranging from He to Na is reconstructed through clustering as seen in Figure 5(d).

{
\setlength{\abovecaptionskip}{6pt}
\setlength{\belowcaptionskip}{-2pt}

\begin{figure}[htbp]
  \centering
  \captionsetup{font=small}   
  \begin{tabular}{@{}c@{\hspace{4em}}c@{}}
    \includegraphics[width=0.40\textwidth,height=0.24\textwidth]{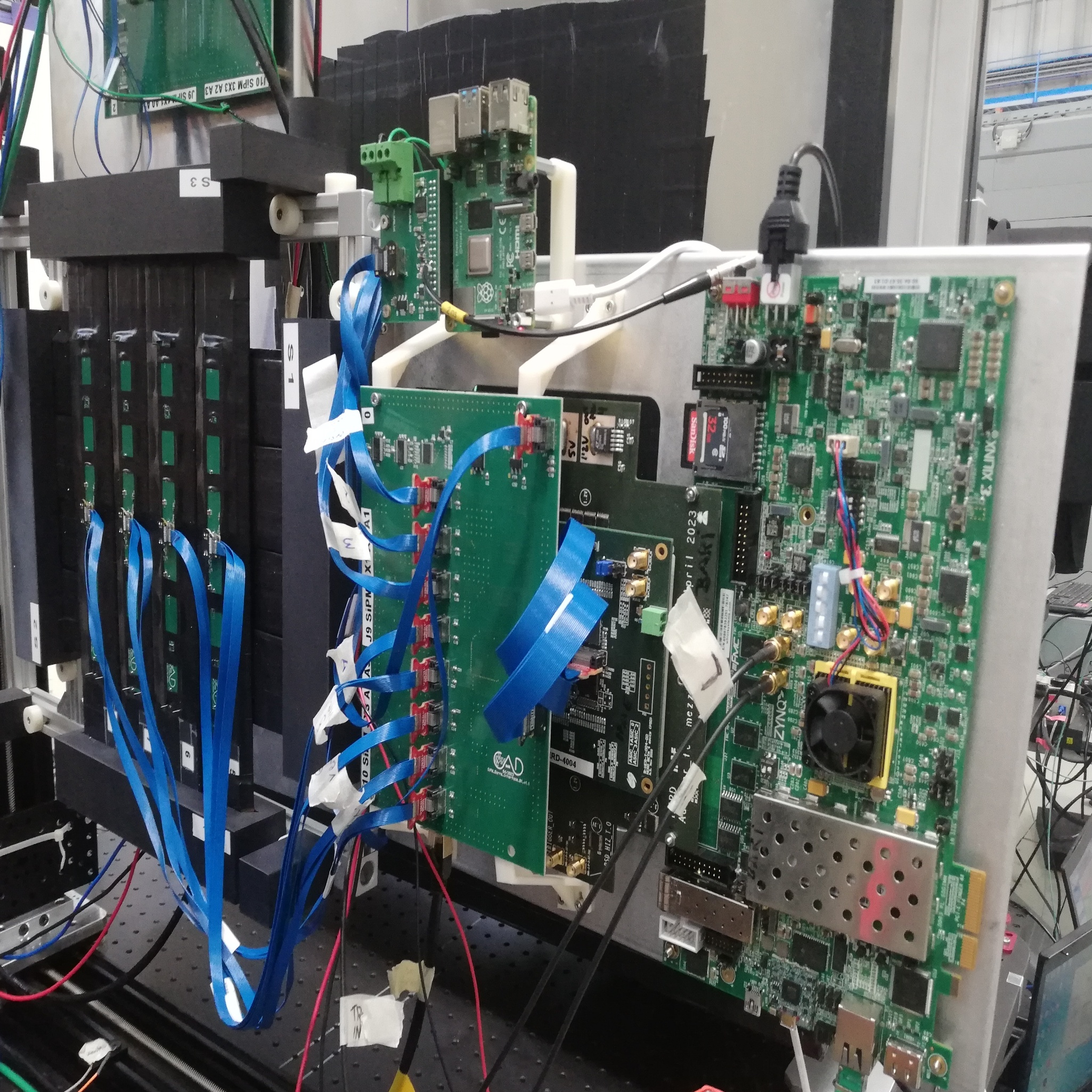} &
    \includegraphics[width=0.58\textwidth,height=0.25\textwidth]{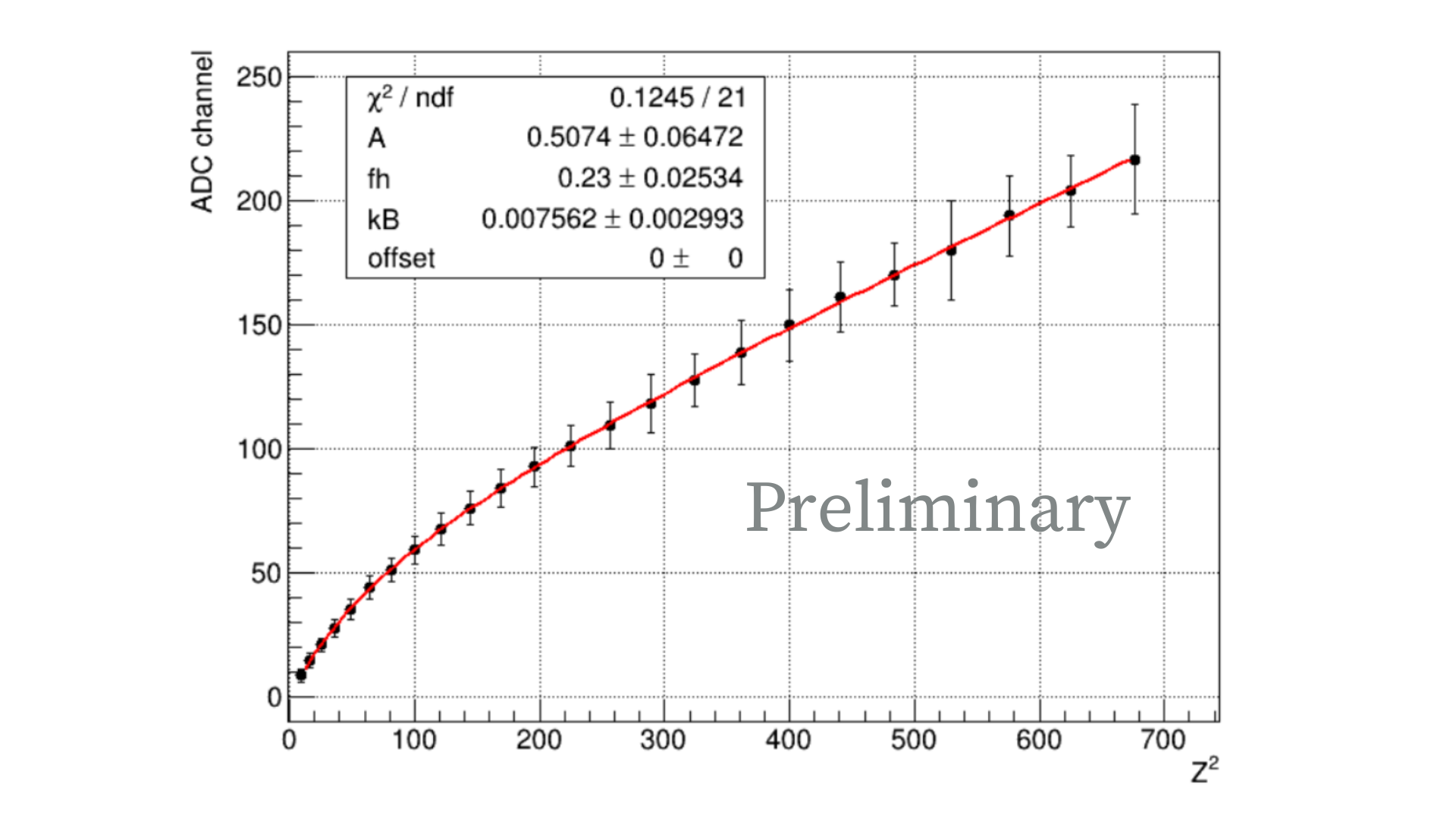} \\ 
  \end{tabular}
  \caption{On the left: picture of the PSD prototype composed of 8 scintillator
           bars tested at CERN PS 2023. On the right: mean signal, measured in
           ADC channels as a function of the square of nuclei charge, $Z^{2}$,
           identified by the PSD. The red line represents the best fit obtained
           using Birk’s function \cite{9}.}
  \label{fig:PSD-panels}
\end{figure}
} 

{
\setlength{\abovecaptionskip}{6pt}
\setlength{\belowcaptionskip}{-2pt}

\captionsetup[subfigure]{%
  font=small,
  labelformat=parens,   
  labelsep=space,
  justification=centering,
  singlelinecheck=false}

\begin{figure}[htbp]
  \centering
  \captionsetup{font=small, justification=justified,
                singlelinecheck=false, width=\textwidth}
  \begin{tabular}{@{}c@{\hspace{7em}}c@{}}
    \subfloat[]{%
      \includegraphics[width=0.40\textwidth,height=0.24\textwidth]{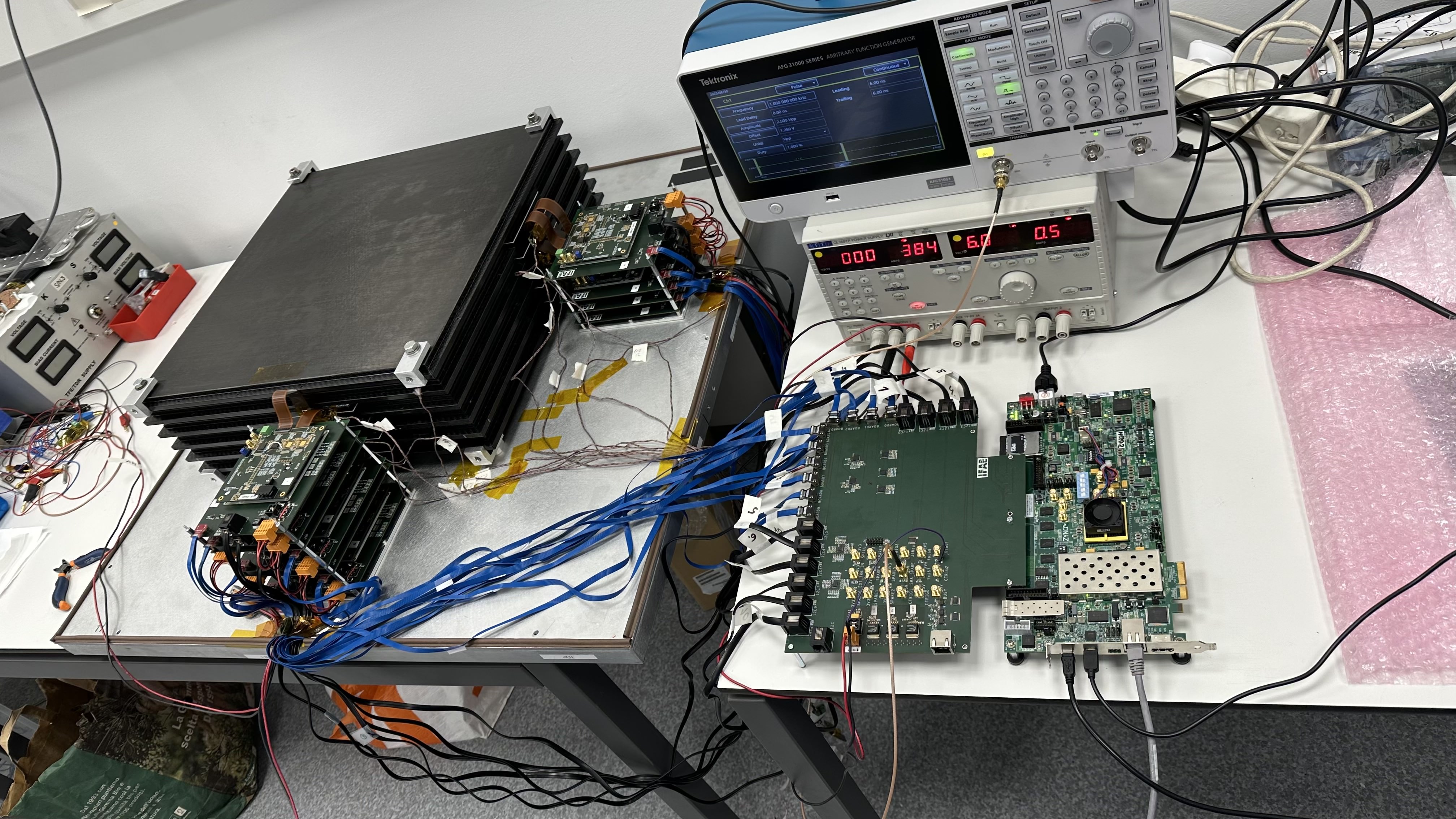}} &
    \subfloat[]{%
      \includegraphics[width=0.35\textwidth]{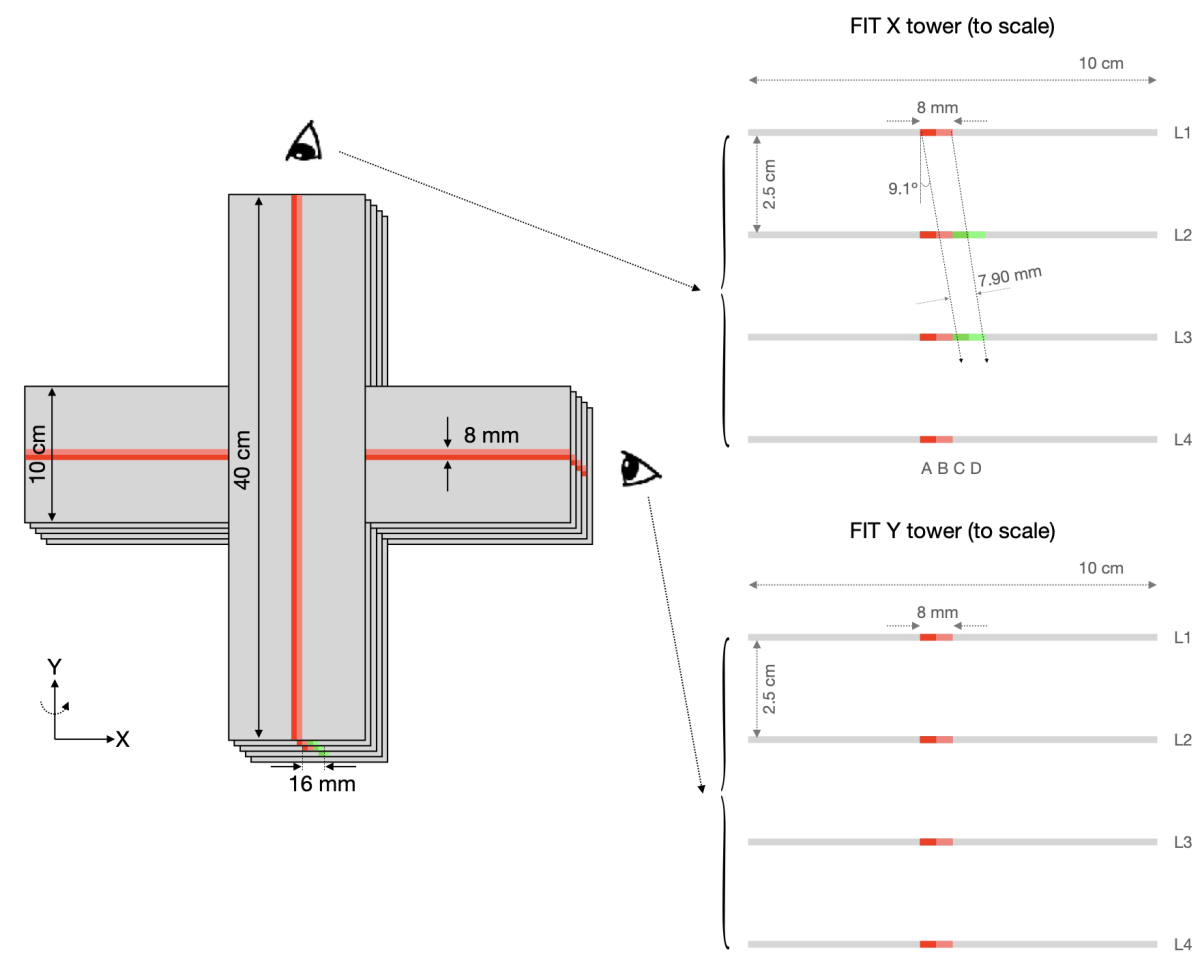}} \\[0.8cm]  

    \subfloat[]{%
      \includegraphics[width=0.40\textwidth,height=0.24\textwidth]{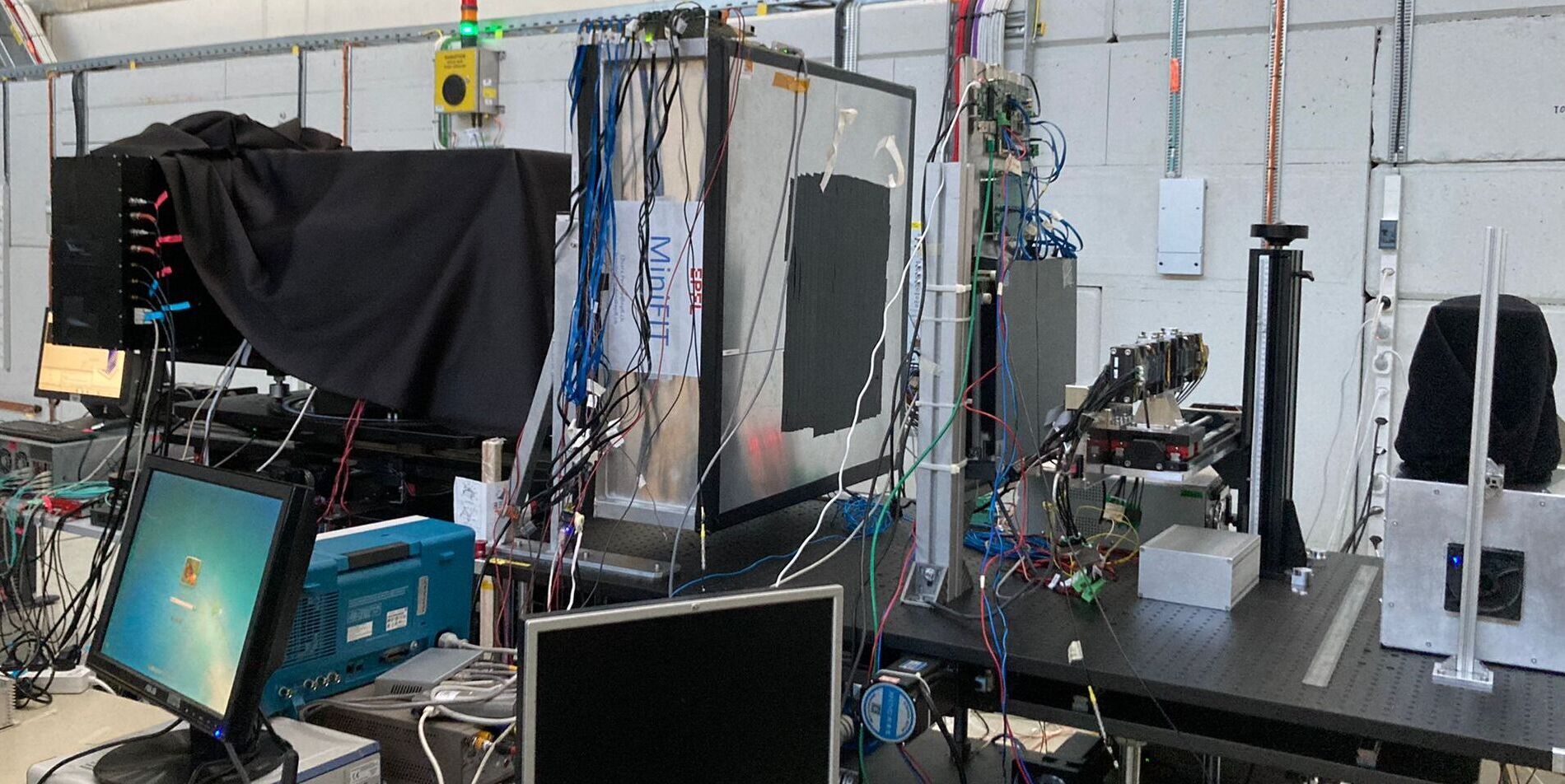}} &
    \subfloat[]{%
      \includegraphics[width=0.42\textwidth,height=0.24\textwidth]{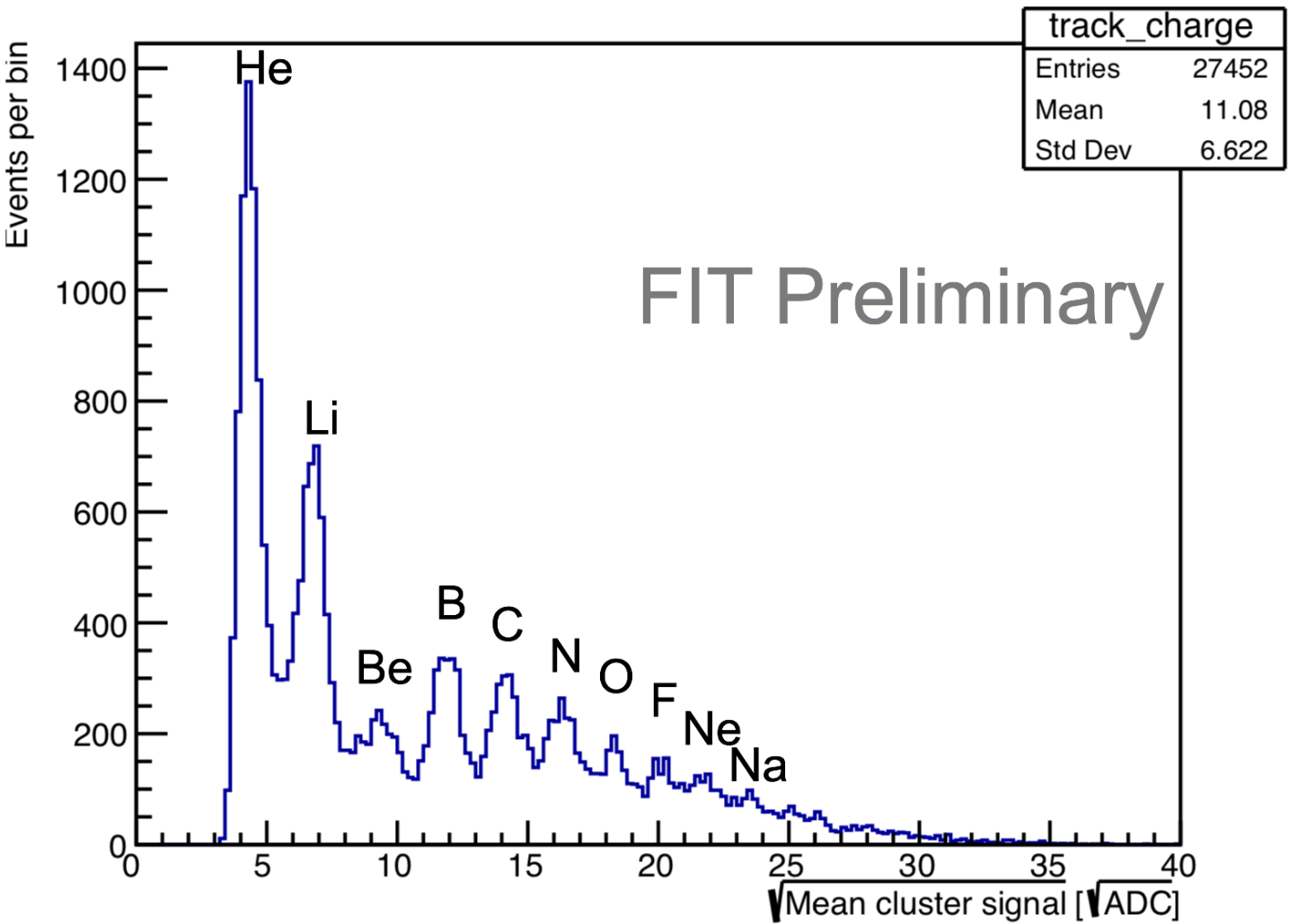}} \\  
  \end{tabular}

  \caption{(a) Picture of the MiniFIT prototype integrated at the University of
           Geneva (b) Schematics of instrumentation of MiniFIT (c) MiniFIT
           inside a light‑tight box tested at CERN PS 2023 (d) Mean signal,
           measured in ADC channels, as a function of the square root of the
           average signal of particles incident in the X and Y towers of
           MiniFIT.}
  \label{fig:MiniFIT-grid}
\end{figure}
}

\section{Conclusion}
We have presented the performance of the ULEG trigger for HERD and the results from the beam tests, which were aimed at validating the hardware implementation of the ULEG trigger in the FIT and PSD prototypes. Currently, the efforts are focused on calibrating and optimizing the hardware, with a specific emphasis on SiPM calibration and optimization using LED pulses. Figure 6 shows the signal distribution across varying LED intensities for a channel of SiPM readout.  In the future, a new version of Beta ASIC with 64 channels for FIT will be used for the beam test 2024, and a PSD model that closely resembles the flight model will be developed.

\begin{figure}[h]
    \centering
    \includegraphics[width=0.8\linewidth, height=6.7cm]{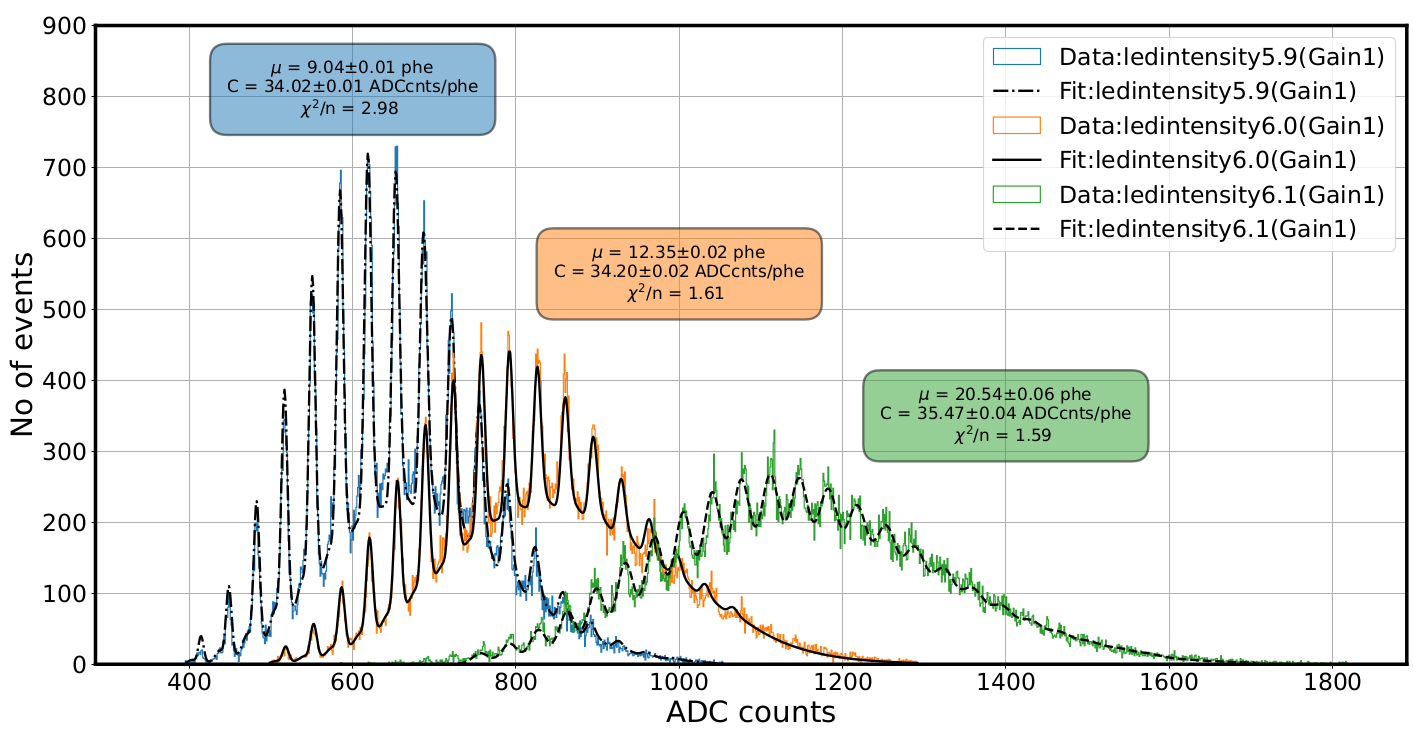}
    \captionsetup{justification=justified, singlelinecheck=false, width=\textwidth}
    \caption{Photoelectron spectrum of a single SiPM channel, recorded using the Beta ASIC chip, under varying LED intensities. The black curves represents the fitted model for each spectrum. Here, $\mu$ denotes the mean number of photoelectrons in the signal distribution, and $C$ represent the calibration factor, which is the difference in ADC counts between every two consecutive photoelectron peaks.}
\end{figure}

\section{Acknowledgements}

This work is supported by Agencia Estatal de Investigaci\'on, Ministerio de Ciencia e Investigaci\'on: PID2020-116075GB-C22/AEI/10.13039/501100011033, PRE2019-091232, PRE2021-097518; European Union NextGenerationEU: PRTR-C17.I1; Generalitat de Catalunya; Swiss National Science Foundation (SNSF-PZ00P2\_193523).

\end{document}